# New insights in gill/buccal rhythm spiking activity and CO2 sensitivity in pre- and postmetamorphic tadpoles (*Pelophylax ridibundus*)

Brigitte QUENET[a], Christian STRAUS [b, c],
Marie-Noëlle FIAMMA[b], Isabelle RIVALS [a],
Thomas SIMILOWSKI [b, d], Ginette HORCHOLLE-BOSSAVIT [a]

a. ESPCI-ParisTech, Equipe de Statistique Appliquée, F-75005 Paris, France

b. UPMC Univ Paris 06, ER 10 UPMC, F-75013 Paris, France

c. Assistance Publique - Hôpitaux de Paris, Groupe Hospitalier Pitié-Salpêtrière, Service Central d'Explorations Fonctionnelles Respiratoires, F-75013 Paris, France

d. Assistance Publique - Hôpitaux de Paris, Groupe Hospitalier Pitié-Salpêtrière, Service de Pneumologie et Réanimation Médicale, F-75013 Paris, France

Corresponding author at: ESPCI-ParisTech, Equipe de Statistique Appliquée, F-75005 Paris, France. Tel.: +33 140794461; Fax: +33 1 40794669 E-mail address Brigitte.quenet@espci.fr (B Quenet).

## ABSTRACT
Central $CO_2$ chemosensitivity is crucial for all air-breathing vertebrates and raises the question of its role in ventilatory rhythmogenesis. In this study, neurograms of ventilatory motor outputs recorded in facial nerve of premetamorphic and postmetamorphic tadpole isolated brainstems, under normo- and hypercapnia, are investigated using Continuous Wavelet Transform spectral analysis for buccal activity and computation of number and amplitude of spikes during buccal and lung activities. Buccal bursts exhibit fast oscillations (20-30Hz) that are prominent in premetamorphic tadpoles: they result from the presence in periodic time windows of high amplitude spikes. Hypercapnia systematically decreases the frequency of buccal rhythm in both pre- and postmetamorphic tadpoles, by a lengthening of the interburst duration. In postmetamorphic tadpoles, hypercapnia reduces buccal burst amplitude and unmasks small fast oscillations. Our results suggest a common effect of the hypercapnia on the buccal part of the Central Pattern Generator in all tadpoles and a possible effect at the level of the motoneuron recruitment in postmetamorphic tadpoles.

# 1. Introduction

Previous studies of the breathing control in amphibians, characterized by a developmental transition from water breathing to air breathing, have addressed the issue of the emergence of central $CO_2$ chemosensitivity. Many investigations use the fact that cranial nerve outputs of the isolated brainstem preparations from tadpoles and frogs match the corresponding respiratory muscle sequences (Gdovin et al., 1998; Kogo et al., 1994a, 1994b; Sakakibara, 1984a, 1984b). The nerve activity exhibits regular low amplitude bursts associated with gill ventilation and episodic high amplitude bursts corresponding to lung ventilation (Galante et al., 1996).

Several studies have shown that hypercapnia increases systematically the occurrence of lung episodes in postmetamorphic tadpoles (Kinkead and Milsom, 1994; Torgerson et al., 1997; Taylor et al., 2003a, 2003b; Wilson et al., 2002; Straus et al., 2011). But regarding the buccal rhythm, experimental results describe heterogeneous effects of hypercapnia and metamorphosis on the gill/buccal frequency in pre- and postmetamorphic tadpoles, leading to contradictory conclusions. In *Lithobates catesbeianus* (*Rana catesbaiana*) of the same premetamorphic stages exposed to hypercapnia, Torgerson et al. (1997) reported an increase in the frequency and the amplitude of the gill rhythm while Taylor et al. (2003a) observed a tendency for frequency to decrease and no change in duty cycle. In postmetamorphic animals of the same stage exposed to hypercapnia, Torgerson et al. (1997) did not observe any response; Taylor et al. (2003b) observed a tendency for frequency to decrease. These conflicting conclusions could be explained by physiological differences between species or/and by methodological differences. Several studies reported effect of metamorphosis at normocapnia in *Lithobates catesbeianus* (Burggren and Doyle, 1986; Togerson et al., 1998; Gdovin et al., 1999; Taylor et al., 2003b; Hedrick et al., 2005; Fournier and Kinkead, 2006; Chen and Hedrick, 2008). These studies gave conflicting results about the gill/buccal activity: for instance, Taylor et al. (2003b) and Fournier and Kinkead (2006) reported no change of the buccal frequency, while Chen and Hedrick (2008) observed a decrease of about 50%.

Developmental observations in tadpoles (Torgerson et al., 2001a, 2001b; Taylor et al., 2003b) have suggested that the neural substrates for central chemoreception exist in the vicinity of the structures responsible for both gill/buccal and lung rhythm generation already in premetamorphic tadpoles. Here, we study how $CO_2$ sensitivity affects premetamorphic and postmetamorphic gill/buccal and lung rhythm in *Pelophylax ridibundus*. We analyze neurograms recorded from isolated brainstems in *Pelophylax ridibundus* whose chaotic nature has been evidenced by Straus et al. (2011). On the part of the filtered neurograms corresponding to buccal activity only, we perform time-frequency transform with complex Morlet wavelets (CWT) that is adapted to non-stationary signals (Akay, 2005; Mor and Lev-Tov, 2007; Vialatte et al., 2007; Romcy-Pereira et al., 2008). With adapted signal-processing tools, we evidenced, for the first time in amphibians, the presence of fast oscillations in the 20-30 Hz frequency band in the buccal bursts, probably resulting from phase locked motoneuron activities, which are reminiscent of some of the fast oscillations described in several species of mammals (Funk and Parkis, 2002; Marchenko and Rogers, 2006, 2007). Regarding the effects of hypercapnia (acidosis at pH=7.4) on the gill/buccal activity, our results show that there is no change of the fast oscillations in the premetamorphic tadpole, but significant changes in postmetamorphic ones. In addition, hypercapnia induces significant lengthening of the buccal interburst period at both stages.

## 2. Materials and methods

*2.1. Data collection*

Using the brainstem preparation of the tadpole, *Pelophylax ridibundus (Rana esculenta* - European Edible Frog) described in detail in Straus et al. (2011), the present study investigates a subset of the signals used in this paper. The experimental protocol fulfilled appropriate legal and ethical conditions (French Ministry for Agriculture and Animal Care Committee - Ile-de-France, Paris, Comité 3). Briefly, the developmental stage of the animals is determined according to Taylor and Köllros (1946). Tadpoles are chosen in order to assign them either to the premetamorphic (stages X-XIII) or to the postmetamorphic (stage XXIV-XXV) group of development. They are anesthetized and decerebrated. The brainstem is dissected and transferred to a recording chamber. This chamber is superfused at room temperature (~21°C) with mock cerebrospinal fluid (CSF) equilibrated with a gas mixture of $O_2$ and $CO_2$: (in mM) NaCl, 104; KCl, 4; MgCl2, 1.4; D-glucose, 10; $NaHCO_3$, 25; $CaCl_2$ 2.4. The pH of the CSF is set to 7.8 or 7.4 by adjusting the fractional concentration of $CO_2$. The electrical activity – namely the neurogram - of the root of the facial nerve (VII) is recorded using glass suction electrodes. The raw signal is then amplified (10 000 times) and filtered (100 Hz–1 kHz) using a high gain, differential AC amplifier (model 1700, A-M Systems Inc., Evrett, WA), digitized at 2000 Hz (PowerLab, ADInstruments) and stored as text files. Recordings last 20-35 min for each pH value. The first 10 minutes of each recording is discarded. Then, 5 minutes of signal with the best identifiable bursts of buccal and lung ventilation are selected for analysis. Indeed, the tadpole isolated brainstem preparation exhibits neural ventilatory activity but also periods of apnea and, sometimes, also bursts of undetermined kind, like, for example, the so-called 'complex discharges' (Straus et al., 2000). We retain the part of the neurograms where the preparation is not apneic at that time the type of the bursts is clear, either buccal or lung, with a minimum number of discharges of an undeterminable kind.

*2.2. Neurogram processing*

The analysis is performed on neurograms from 5 premetamorphic (T, 1-5) and 5 postmetamorphic (F, 1-5) preparations, in normo- (pH=7.8) and hypercapnia (acidosis at pH=7.4). The spectral analysis based on complex Morlet wavelets (CWT) leads to information in the frequency domain taking temporal short-time variations into account. After segmentation of the signal into cycles (buccal and lung), we identified the lung bursts, and computed cross-correlation functions in order to insure the best temporal alignment of these segmented cycles of each type, buccal or lung, separately. Finally, we come back to the neurograms, which can be analyzed using the previous segmentation and temporal alignment of the cycles, in order to measure the position and amplitude of each spike in each cycle.

*2.3. Neurogram filtering and spike detection*

Each neurogram is squared, zero-phase filtered (Gustafsson, 1996) using a moving average rectangular window of width $W=100$ ms. After taking the square root of the result, the lung bursts are eliminated (see next paragraph). On the resulting $S_{100}$ signal, a detection of local minima is performed in order to segment this buccal signal into cycles, preserving both the buccal burst and the inactive period in each cycle (Fig. 1A, black line). The same filter with a window of width $W=10$ ms is applied to the buccal part of the neurogram, $S_{neur}$ (Fig. 1A, gray line) resulting in a new signal $S_{10}$ (Fig. 1A, red line).

A detection of maxima, i.e. spikes, is performed independently on the positive and the negative parts, in absolute value, on $S_{neur}$ (Fig. 1B, red and blue dots), in order to store the time position and the amplitude of each of them. This detection allows the computation of a local spike density (per time unit).

*2.4. Lung burst detection and analysis*

The detection of local minima performed on the squared and filtered neurogram (with $W$=100 ms) is completed by a detection of each local maximum between two successive minima defining each burst. In a given neurogram, a burst is labeled 'lung burst' according to two criteria: its amplitude is higher than twice the mean amplitude of all the recorded burst, and its spike density is higher than twice the mean spike density of all the recorded bursts.

*2.5. Cycle time alignment by cross-correlation function*

The aim of this signal-processing step is to get a time alignment of the buccal bursts able to exhibit phase locking of neuronal activity inside the bursts of a given neurogram. Therefore, we define in $S_{10}$ a "reference cycle" and position all the other cycles according to the maximal value of the cross-correlation function (sliding dot product) between each cycle and the reference (Lynn and Fuerst, 2000). At a first step, the amplitude vectors representing the $S_{10}$ buccal cycles, after segmentation, are padded with zeros when necessary, in order to get an identical length for all these vectors. In a second step, a symmetric matrix of correlations is computed, whose generic term of indices $i$ and $j$ is the maximal value of the cross-correlation function between the amplitude vectors of cycle $i$ and cycle $j$. The "reference cycle" is then the cycle whose mean correlation with all the other cycles is the highest: it is therefore considered as the most representative cycle. At a third step, the time lag, which corresponds to the maximum of the cross-correlation function between a cycle and the reference cycle, defines the time position of this cycle relatively to the reference (Fig. 2A): when the cycles are positioned according to their time lag and ranked according to their correlation with the reference, the bursts are aligned, and each cycle can be visualized in a map by a horizontal band where its amplitude is coded in gray scale (Fig. 2A). A same time alignment procedure is applied to the lung burst.

*2.6. Amplitude profiles from the "spike matrices"*

The aim of this procedure is to get canonical profiles of the burst and interburst part representative of each neurogram buccal cycles. Using the relative time positions defined previously, we build two matrices of the cycle positive and negative spikes respectively (Fig 1B). For each matrix, the generic term of indices $i$ and $j$ is the amplitude of a spike, when present, at sampling time $j$ of cycle $i$. In maps visualizing these matrices, each row corresponds to a cycle, each column corresponds to a time bin of the cycles, and a dot in a row and a column corresponds to a spike in that time bin for this cycle (Fig 2B). An amplitude profile is computed from these two matrices: the mean values, at each time bin, of the positive and negative amplitudes over the cycles, after zero-phase filtering with a window of 5 ms, constitute the two parts of the profile.
In order to define a canonical cycle for each neurogram, characterized by its amplitude profile, with its positive and negative parts, whose duration corresponds to the median value of the cycle durations, we use the most representative part of the amplitude profiles, i.e. where the number of zeros related to the padding is the smallest: this criterion defines the beginning and the end of the canonical cycle (Fig. 2A and 2B). In such a canonical cycle, we want to define a central active zone, or duty cycle, corresponding to the buccal burst, surrounded by zones of minor activity. The beginning of the burst is the first time when the positive amplitude profile is higher than its mean value; symmetrically, the end of the burst is the last time when the positive amplitude profile is higher that its mean value. As the positive and the negative amplitude profiles are very similar in absolute value, only the positive part is visualized when comparing the canonical cycles at the two values of the pH; moreover, in graphical representations of a canonical activity, the positive part of the canonical cycle is repeated twice, in order to get a complete view of the interburst part. Amplitude profiles of lung burst are computed using the same procedure.

## 2.7. Number and amplitude profiles according to spike amplitude

The aim of this procedure is to evaluate the contributions of the spikes to the time structure of the bursts according to their amplitude. The alternation of high and weak spike activity characterizes the cyclic buccal part of the neurogram: we define the high activity window (buccal burst duration or 'duty cycle') as the time interval of the buccal canonical cycle where its profile amplitude is higher than the mean amplitude of this cycle (between the vertical green lines in Fig. 2B). Then, two histograms of spike amplitude are computed in the buccal cycles -aligned according to the previously described cross-correlation procedure (see § 2.5)- corresponding to the spikes inside the active window ($P_{in}$) and to the spikes outside ($P_{out}$) this window respectively. Populations $P_{in}$ and $P_{out}$ exhibit lognormal-like shaped distributions (not shown) of the spike amplitudes. The threshold amplitude value $A_{th}$ where the number of spikes of $P_{in}$ becomes higher that the number of spikes of $P_{out}$, defines the minimal amplitude of a spike to have a higher probability to belong to $P_{in}$ than to $P_{out}$. The positive spike population is then divided into five subpopulations. The first one corresponds to the spikes whose amplitude is smaller than $A_{th}$, the second, third and fourth subpopulations contain spikes whose amplitude belongs to $]A_{th}, 3*A_{th}]$, $]3*A_{th}, 5*A_{th}]$ and $]5*A_{th}, 7*A_{th}]$ respectively, and the fifth subpopulation contains the highest remaining spikes, higher than $7*A_t$ (Fig. 1B). For each subpopulation $k$ ($k$=1, 2, 3, 4 or 5), two new matrices are computed: a matrix of the number of spikes over ten consecutive time bins $M^k$ and a matrix of the mean amplitude of these spikes $P^k$. In order to visualize the time distribution of the spiking activity for each subpopulation, $M^k$ is graphically represented by a map of dots, a dot corresponding to the presence of one spike at least (Fig. 2B). The time successive mean values of the number of spikes over the cycles define a number profile for each subpopulation, idem for the time successive mean values of the amplitude, which define the amplitude profile for each subpopulation.

## 2.8. Continuous Wavelet Transform maps and spectra

Given the non-stationarity of the buccal rhythm, a time-frequency analysis of $S_{10}$ signal is performed using a Continuous Wavelet Transform (CWT) (Akay, 2005; Mor & Lev-Tov; 2007; Vialatte et al., 2007; Romcy-Pereira et al., 2008) of $S_{10}$ subsampled at 200 Hz (after applying an anti-aliasing low pass filter eliminating all the frequencies above 100 Hz), looking for relevant frequencies between 1 and 100 Hz, with a complex Morlet wavelet basis function including five oscillations (Mallat, 2000). The coefficients of each time-frequency amplitude map are computed for 100 values logarithmically distributed between 1 and 100 Hz, and at a time resolution corresponding to the sampling of $S_{10}$. This time-frequency map displays the temporal evolution of the $S_{10}$ dominant frequencies in gray scale (Fig. 3). All the CWT maps got from our signals exhibit a relative stability of the dominant frequencies, which validates the representation of each $S_{10}$ by a CWT spectrum computed as the mean value, at each frequency, of the corresponding coefficients of the amplitude map over time activity (Fig. 3). In order to compare these spectra, each of them is normalized.

*2.9. Statistical analysis*

Statistical tests have been applied to the three following descriptors: the mean buccal cycle duration, the relative energy in the band 1-10 Hz and the relative energy in the band 20-30 Hz, measured on the CWT spectrum. In order to test both the effects of pH and stage on these descriptors, a two-way ANOVA is applied when the condition of homoscedasticity is fulfilled, which is tested with Bartlett's test. In this case, we also have to take into account the fact that the same animal is tested for the two values of the pH, by including a random individual effect. When homoscedasticity is rejected, pairwise statistical tests are performed. The effect of the pH at a given stage is tested with a paired t-test, the effect of the stage at a given pH is tested with a standard t-test if the variance of the two groups are homogeneous (F-test), with a Wilcoxon rank sum test if not. Statistical significance was assumed if $p < 0.05$. Signal processing algorithms and statistical computations have been implemented with Matlab (The Mathworks, version 8.0.0.783 (R2012b)).

# 3. Results

*3.1. CWT maps and spectra of buccal activity*

Frequency information from $S_{10}$ signal is visualized in both CWT maps and spectra, as illustrated for T2 and F4 at both values of the pH (Fig. 3, lower part). Three seconds of the corresponding $S_{10}$ signals are illustrated in the upper part of this figure. In time domain, the main effects induced by metamorphosis and by acidosis on the buccal rhythm are already visible, i.e. a shortening of the buccal period after metamorphosis, a lengthening of the period with acidosis in both premetamorphic and postmetamorphic tadpoles, and oscillations superimposed on the buccal bursts at both pH in premetamorphic tadpoles, much more discrete in postmetamorphic tadpoles.

For T2, the maps reveal a dark gray strip, discontinuous, at low frequency, 1.5 Hz, corresponding to the fundamental buccal frequency at pH=7.8, whereas the equivalent strip is darker, continuous, at 1.2 Hz at pH=7.4, which indicates that hypercapnia induces a buccal activity at lower frequency and at higher rhythm regularity. A wider strip, at high frequency centered at 21 Hz, is present with same characteristics at both values of the pH, corresponding to fast oscillations exhibited on the bursts in $S_{10}$ signal. For F4, the maps reveal a dark, continuous, low frequency strip (3.7 Hz) at pH=7.8, while the corresponding strip at pH=7.4 is lighter, at 2.5 Hz. Here, hypercapnia induces a lower and more variable frequency of the buccal activity. A second strip is present at both pH values, at about twice the corresponding low frequencies.

All these pieces of frequency information are also given in the spectra, which are illustrated for all the signals in Fig. 4. For each premetamorphic tadpole at pH=7.8 and 7.4, the spectra reveal two or three local maxima: the first one occurs systematically in the low frequency range [1.1, 1.6] Hz, and the last one occurs in the high frequency range [20.6, 27.2] Hz. The frequency of the first peak, or buccal low frequency, is slightly lower in hypercapnia than in normocapnia and the amplitude of this peak is higher, but for T5. The high frequency parts of the spectra are very similar at both pH.

In the postmetamorphic tadpole, three maxima can be observed in the spectra. In normocapnia, a first peak occurs at low frequency in the range [2.9, 3.7] Hz, systematically followed by the peak at twice this fundamental frequency, i.e. [5.9 7.4] Hz. In two cases (F1, F5) there is a smooth local maximum around 22 Hz. In hypercapnia, the first peak occurs at a lower frequency, in the range [1.5, 2.5] Hz. When present, the second peak is very small in amplitude and a smooth local maximum at high frequency around 22 Hz can be seen in all cases. In contrast with the premetamorphic tadpoles spectra, the amplitude of the first peak is systematically smaller in hypercapnia than in normocapnia, while there is an increase in the high frequency band.

The statistical tests performed on the CWT spectral energies in the low frequency (1-10 Hz) and in the high frequency band (20-30 Hz) give the following results. For the effect of metamorphosis, the relative energy in the low frequency band is significantly higher for the postmetamorphic tadpoles at both pH: at pH=7.8, +0.33 (p=0.008) and at pH=7.4, +0.09 (p=0.029), while in the high frequency band it is significantly lower: at pH=7.8, -0.12 (p=0.008) and at pH=7.4, -0.04 (p=0.001).

For the effect of the acidosis in premetamorphic tadpoles, there is no significant effect on the relative spectral energy neither in the low frequency band nor in the high frequency band. In postmetamorphic tadpoles, the decrease in the low frequency band is significant: -0.23 (p=0.023), as well as the increase in the high frequency band: +0.07 (p=0.012).

### 3.2. Amplitude profiles of the buccal canonical cycles

In order to visualize in the temporal domain the dynamical modifications induced in the buccal rhythm by hypercapnia, we superimpose the canonical activities at both pH. A graphical comparison is thus obtained by positioning the positive part of the amplitude profile of the canonical activity (i.e. a canonical cycle repeated twice) at pH=7.4 in such a way that the correlation of its first cycle with the first cycle of the canonical activity at pH=7.8 is maximal.

These amplitude profiles show that oscillations are systematically present in premetamorphic tadpoles in normo and hypercapnia. For all premetamorphic tadpoles, oscillations occupy the major part of the cycle (Fig. 5, left). These oscillations are rather regular, with an interval of about 35-50 ms between two successive peaks. Under hypercapnia, the cycle duration is systematically increased by about 0.14 s, i.e. by around 20%. This duration increase clearly affects the interburst part of the cycle. The burst amplitude and duration seem very weakly modified.

For postmetamorphic tadpoles, the bursts are shorter and exhibit an asymmetry, with a sudden interruption of activity which ends the burst, with very few and small oscillations, if any (Fig. 5, right). Under hypercapnia, the cycle duration is systematically increased by about 0.18 s, i.e. by around 60%. This duration increase clearly affects the interburst part of the cycle also. The burst amplitude is markedly reduced with visible intra-burst oscillations.

The N-way ANOVA performed on the mean value of the buccal period (Table 1) indicates that there is no individual effect, but a significant stage effect with a decrease of 335 ms (p=0) and a significant pH effect with an increase of 155 ms (p=0.0004), and no significant cross-effect between pH and stage, which suggests a similar lengthening of the period with the pH for both pre- and postmetamorphic tadpoles.

### 3.3. Spike contributions to the buccal activity according to their amplitude

For T2, the matrices $M^k$ containing the number of spikes in ten successive time bins (i.e. 5 ms) for each of the first four subpopulations (k=1,2,3 and 4) are visualized as maps of gray dots for pH=7.8 and red dots for pH=7.4 (Fig. 6). By definition, the first subpopulation contains the smallest 'spikes', with both neuron firings and noise, whose density is higher outside the burst than inside (the burst limits are represented by the green lines in Fig.6): this outside density is greater in hypercapnia than in normocapnia, while the inside burst density is quite the same at both pH values. These features characterize both the number and amplitude profiles, even if this high spike density brings a small contribution to the amplitude profile (Fig. 6). For the second subpopulation, there is a small increase in spike density and amplitude during the burst with slightly smaller values at pH=7.4, particularly during the prolonged interburst part of the cycle. For the third population, there is an increase in spike density and amplitude during the burst, carrying out its plateau part, with a same activity profile of the burst at both pH values. The corresponding spike density has a maximum at 0.31 spikes/5ms = 62 spikes/s. For the fourth subpopulation containing the highest spikes, at both pH values we observe the quasi-absence of spikes outside the burst and an alternation of bands of high and low density of these spikes, resulting to the oscillations of the number and amplitude profiles. In conclusion, in T2 and all premetamorphic tadpoles, in addition to the lengthening of the inactive period, the only visible effect of hypercapnia is the reduction of the second subpopulation spikes during the prolonged interburst.

For F2, the matrices $M^k$ (k=1,2,3 and 4) are visualized for pH=7.8 and pH=7.4 in Fig. 7. At pH=7.8, the first subpopulation exhibits a band of very high and constant density at the end of the burst that cannot be of physiological origin: it means that this first peaks/spikes subpopulation is clearly dominated by instrumental noise. For the second population, the spikes are concentrated inside the burst with same densities inside and outside the burst at both pH values, but the duration of the burst is longer at pH=7.4. The spike density of the third subpopulation exhibits an increase during the burst followed by a quasi-absence of spikes after the burst. The spike density inside the burst is smaller in hypercapnia than in normocapnia. These features are accentuated in the fourth subpopulation, where some oscillations are present. For all neurograms from the postmetamorphic tadpoles, as in F2, in addition to the lengthening interburst duration, the main effect of hypercapnia is the strong reduction of the third and fourth subpopulation spikes during the burst.

*3.4. Analysis of the lung activity*

Figure 8 illustrates the intensity of the neuronal activity for the ten neurograms of the postmetamorphic tadpoles represented by the whole corresponding $S_{100}$ signals, with both buccal and lung components. This figure points out the higher amplitude of the buccal bursts at pH = 7.8 than at pH=7.4. The irregular occurrence of lung bursts is visible in the distribution of the vertical bars indicating these lung bursts. The number of lung bursts is higher at pH=7.4 than at pH=7.8 (but F1, whose number of lung bursts is already high at pH=7.8). The left part of figure 9 illustrates two typical lung bursts observed in the neurograms of the postmetamorphic tadpole F2 at pH=7.8 and pH=7.4 respectively: in this tadpole, all the lung bursts at pH=7.8 are preceded by a small amplitude burst, which is totally absent at pH=7.4. The corresponding mean profiles confirm the presence of these two lung burst types; in addition they show that the mean amplitude of the lung bursts is smaller at pH=7.4 than at pH=7.8. Regarding the contribution to the different spike populations according to their amplitude, the main observation is that the lung burst amplitude is related to the spikes of population V that appears to be specific of the lung activity and seems to be more present in normocapnia than in hypercapnia in F2. Of note, there is no visible temporal structure in the lung bursts as observed in the buccal bursts.

# 4. DISCUSSION

In the present paper, our analysis of the neurograms was done in order to capture some dynamic characteristics of buccal and lung rhythm generator at short-time scale and to evaluate the effects of hypercapnia on these characteristics. In the European frog, *Pelophylax ridibundus* the major findings of the study are the following:
1. In normocapnia, the frequency of the buccal rhythm in premetamorphic tadpoles is in the range of [75, 96] min-1 and in postmetamorphic tadpoles, in the range [151, 220] min-1, which is higher than all reported values in the American bullfrog (Table 1).

2. Fast oscillations in the frequency band 20-30 Hz are present in all tadpoles, revealed by both the spectral analysis and by the buccal amplitude profiles (Fig. 4 and Fig. 5). Such oscillations are visible in other amphibian species under the form of small indentations in the standard filtered neurograms, but they have not been studied.
3. The analysis of spike time positions and amplitudes reveals their respective contributions to the burst structure: the small spikes are uniformly distributed in the burst, while the high ones are present in specific time windows separated by about 35-50 ms at the origin of the fast oscillations (Fig. 6 and Fig. 7).
4. Metamorphosis induces a significant shortening of the cycle duration of about 335 ms (Table 1)
5. Metamorphosis induces a significant increase of the relative energy in the low frequency band (1-10 Hz) and a significant decrease of relative energy in the high frequency band (20-30 Hz) (Fig. 4).
6. Hypercapnia induces a significant lengthening of about 155 ms of the interburst duration in branchial/buccal rhythm in both premetamorphic and postmetamorphic tadpoles (Table 1 and Fig. 5).
7. In the buccal activity of postmetamorphic tadpoles, hypercapnia induces a significant decrease of the relative energy in the low frequency band (1-10 Hz) and a significant increase of relative energy in the high frequency band (20-30 Hz) (Fig. 4).
8. In the buccal activity of postmetamorphic tadpoles, hypercapnia reduces noticeably in the number of high amplitude spikes, which is related to the low frequency energy decrease (Fig. 7).
9. Lung bursts are present in all postmetamorphic tadpoles, and, as in other species, are more frequent in hypercapnia than in normocapnia (Fig. 8).
10. The lung bursts are characterized by the presence of spikes of high amplitude (population V)

From our analysis of the neurograms with their buccal and lung activities, it seems that the lung activity is "phase locked" with the buccal cycles in some preparations. In such preparation, the phase locking may be different depending on the pH (for example see type I or type II lung bursts of figure 9). This dynamic relationship suggests reliable dynamical interconnections between the "lung network" and the "buccal network" of the CPG.

Our observations performed on in vitro brainstem preparations of tadpoles of the European frog are similar to those of Taylor et al (2003a and b) with the same type of preparation in *Lithobates catesbeianus*: hypercapnia is associated with a frequency decrease in buccal oscillations in pre- and post-metamorphic tadpoles through a lengthening of the interburst interval. In the case of premetamorphic tadpoles, the lengthening of the interburst duration induced by hypercapnia leads to a prolonged inactivity of nerve VII motoneurons, which can be attributed to a modification of the buccal rhythm genesis, since there is no lung activity. This hypercapnia-related increase of the interburst in the postmetamorphic tadpoles induces a higher effect on the buccal frequency, since this frequency in normocapnia is much higher in postmetamorphic tadpoles (about 3 Hz) than in premetamorphic tadpoles (about 1.25 Hz). In postmetamorphic tadpoles, this frequency effect is associated to a significant amplitude reduction of the buccal bursts. These two effects that express a decrease in the buccal activity could be the result of a global anesthetic effect of $CO_2$; however, such a global effect would also decrease the lung activity, in contrast with the observed increase of the lung frequency. Both amplitude and frequency reduction of the buccal oscillation are also observed in vivo in adult *Lithobates catesbeianus* (Kinkead and Milsom, 1994). In air-breathing postmetamorphic tadpoles, several hypotheses have been proposed for the physiological function of the buccal rhythm, which is no longer directly involved in gas exchange. The buccal rhythm ventilates the oropharynx alone: it has been speculated that is optimizes the efficiency of lung breaths by keeping the buccal cavity ventilated with fresh air (Gans et al., 1969). This point of view has been challenged by the attribution of a likely olfactory function to the buccal activity (Foxon, 1964; Vitalis and Shelton, 1990; Jorgensen, 2000).

The hypercapnia frequency effect seems to be a buccal Central Pattern Generator (CPG) chemosensitive property already present before metamorphosis and retained after metamorphosis despite the loss of a direct ventilatory function of the buccal activity. Our results evidence that the amplitude decrease of the buccal burst induced by hypercapnia in postmetamorphic tadpoles is due to a smaller number of high amplitude spikes. This reduced motoneuronal activity could result either from a chemosensitivity of a motoneuron population whose threshold may be increased at pH=7.4 or from a less efficient excitatory input from the buccal CPG. In both cases, the motor units that are not recruited in hypercapnia during the buccal activity are likely more efficiently recruited during lung episodes, which occur at a higher frequency than in normocapnia.

The CWT spectra of the buccal activity (Fig. 4) reveal the peak at 20-30 Hz (Fig. 4) corresponding to the oscillations observed in $S_{10}$. These oscillations are mainly visible in the premetamorphic tadpole buccal amplitude profiles (Fig. 5). When focusing on the contribution of the spike sub-populations defined by their amplitude, the spike matrices reveal a quasi-periodic density variation of the fourth population spikes (Fig. 6, top and left), with a period of about 40 ms repeated about 10 times during a buccal burst. This density variation of the highest spikes explains the oscillatory feature of the curves representing their mean number and mean amplitude, and finally the complete amplitude profile (Fig. 5). It is quite unlikely that this temporally organized motoneuron recruitment process is due to intrinsic motoneuron properties, thus we suppose that it is a consequence of the driving of the motoneurons by the buccal CPG, which may send oscillating excitatory inputs to the motoneuronal population. If we suppose that, in premetamorphic tadpoles, the motoneuronal population is quite homogeneous, i.e. weakly differentiated, all these motoneurons are driven in the same way by the oscillating excitatory input and exhibit a higher firing probability at the maxima of the oscillation: the presence of high spikes in the corresponding short time windows may thus result from the summation of individual action potentials occurring in coincidence. In postmetamorphic tadpoles, the peak in the CWT spectra at twice the fundamental frequency [5.9 7.4] Hz is prominent in normocapnia (Fig. 4): it accounts for the specific burst shape, higher and narrower than in the premetamorphic tadpoles (Fig. 3 and Fig. 5), that we relate to the recruitment of a more differentiated motoneuron population. Indeed, the asymmetry of the burst shape exhibited in postmetamorphic tadpole amplitude profiles suggests a progressive recruitment of motoneurons related to size effects of a differentiated population, according to the Henneman size principle (Henneman et al., 1965), where the largest neurons, with the highest threshold are recruitment only at the end of the generator buccal cycle. This progressive recruitment is followed by a sudden stop: the fast and complete de-recruitment of the motoneurons at the end of the burst may be due either to an active inhibition process or to the end of the input burst. In normocapnia, such a recruitment process of the motoneurons may also explain the quasi-absence of fast oscillations at pH=7.8, while present at pH=7.4, although these oscillations are likely structuring the buccal generator activity in both cases. In this hypothetical frame, at pH=7.4, the threshold of CO2 chemosensitive motoneurons may be increased in such a way that the generator activity input is not sufficient to recruit them. However, some small motoneurons, with a low threshold, still respond to this oscillating input (Fig. 4).

Fast oscillations have been described in the respiratory neural control system of several mammalian species (for a review see Funk and Parkis, 2002). They are present in inspiratory-related muscles, nerves, and neurons and include both Medium- (MFO) and High-Frequency Oscillations (HFO). The first studies on phrenic and hypoglossal nerve discharges in cats (Richardson and Mitchell, 1982; Cohen et al., 1987; Christakos et al., 1988,1991) revealed frequency peaks in the two bands: 20–50 Hz (MFO) and 60–110 Hz (HFO). Similarly, in adult rats (Marchenko and Rogers, 2006), the time-frequency spectra of phrenic and hypoglossal nerve also exhibited two peaks: one for MFO (37–110 Hz) and one for HFO (156–230 Hz) during eupnea. In mouse (O'Neal et al, .2005), spectral analyses of diaphragm EMG bursts revealed peaks associated with the MFO (20 – 46 Hz) and HFO. In kitten brainstems (Kato et al., 1996), spectra calculated for the inspiratory discharges of facial respiratory nerves showed a stable oscillation in the MFO range only (27-32 Hz) and in newborn rat (Kocsis et al, 1999), this band is also present at (22.8-43.0 Hz). As in kitten, in juvenile rat there is a dominant band in the MFO at 20–55 Hz (Marchenko and Rogers, 2007). Thus, the fast oscillations appear to be dependent on developmental stage, with MFO always present in young animals.

The fast oscillations at 20-30 Hz that we observe in *Pelophylax ridibundus* are reminiscent of MFO in mammals: indeed, Marchenko and Rogers (2007), for instance, attribute the genesis of the MFO band to a central generator that could be of ancestral origin. The presence of fast oscillations in the ventilatory central generator appears as a characteristic that has been retained through the phylogeny, and it would be interesting to look for them in other anuran species, particularly in the well-studied *Lithobates catesbeianus*. This temporal structure of the buccal cycle, in addition to its fundamental low frequency (about 1 Hz) rhythm, suggests the presence of high frequency pacemaker-like activities of the buccal generator, of cell and/or network origin as observed in several types of locomotion, respiration and mastication CPGs (Harris-Warrick, 2010). We suppose that the activity of the buccal network neurons is the consequence of convergence of two pacemaker inputs with two slightly different frequencies on these neurons. The interference between two such inputs may generate both a slow beating frequency (Lefler Y. et al, 2013) (the buccal fundamental rhythm) and a high frequency related to the mean of the pacemaker frequencies (intraburst oscillations), with or without filter effects (Rose G.J. and Fortune E.S., 1999).

In a modeling approach of the buccal and lung CPG, Horcholle-Bossavit and Quenet have proposed (Horcholle-Bossavit and Quenet, 2009) a synchronizing function of the lung part of the CPG in the motoneuron recruitment process, which could account for the high amplitude spikes specific of the lung bursts. The model also implements $CO_2$ sensitivity at the level of the lung part of the CPG to explain the increased lung episodic response to hypercapnia. The results of the present work involve new mechanisms which must be included in the model of the CPG, introducing interactions between pacemakers, in order to implement the buccal $CO_2$ chemosensitivy effects, the fast oscillations structuring the buccal cycle, and the phase-locking of the lung bursts with the buccal activity.

## 5. CONCLUSION

Our results about the buccal ventilatory rhythm control in *Pelophylax ridibundus* evidence the presence of fast oscillations inside each buccal burst due to coincident firings of motoneurons in quasi-periodic time windows. The effects of hypercapnia on the frequency components, i.e. the low frequency (buccal rhythm) and the high frequency (fast oscillations), suggest that chemosensitivity affects both buccal generator elements and some facial motoneurons properties.

## ACKNOWLEGMENTS

We thank Quentin Michard for technical assistance in the time-frequency analysis and Didier Cassereau and Yacine Oussar for methodological advices.

## FIGURE LEGENDS
**Figure 1: Neurogram analyses**
A) By filtering neurogram $S_{neur}$ (gray line), two signals are extracted: $S_{100}$ (black line) and $S_{10}$ (red line). The neurogram and the signals are illustrated for a premetamorphic tadpole neurogram portion during 1.3 seconds that includes two successive buccal cycles, separated at the level of the local minima of $S_{100}$ (blue diamonds). Thus, each neurogram is segmented into successive cycles (red dashed vertical lines).
B) Identification of the spikes by location of the local maxima on the positive (red stars) and negative (blue stars) parts of the neurogram (gray line, same portion as in A). The green lines define five spike subpopulations (I to V) according to their amplitude

**Figure 2: Cycle alignment and burst limit definition**
A) The $S_{10}$ cycle map is built by positioning and stacking cycles using cross-correlation with a reference cycle (red line). In this map, the cycle amplitude values are coded in gray scale. The blue line represents the proportion of zero padding in the map, between 0 and 1.
B) Using the previous positioning of the cycles, spike maps (positive and negative) are built from the spike identification (Fig. 1). The spike amplitude values are represented in gray scale. Positive and negative mean profiles are computed from the respective spike matrices (gray lines); canonical positive and negative cycle amplitude profiles are defined by filtering the previous mean profiles and by cutting these profiles at their most significant portion to define a mean cycle, i.e. where the zero-padding proportion is minimal (between blue lines). The buccal burst is defined as the most active zone of the cycle (between green lines).

**Figure 3: CWT Time-frequency maps and spectra from $S_{10}$ signal**
Upper part: $S_{10}$ signal during 3 seconds for a premetamorphic tadpole (T2) and a postmetamorphic tadpole (F4) in normocapnia (black line) and hypercapnia (red line) with oscillations on the buccal bursts, particularly visible for T2.
Lower part: On the CWT time-frequency maps (shown here for T2 and F4 at pH=7.8 and pH=7.4) of $S_{10}$ signals with a logarithmic distribution of frequencies in the range [1-100] Hz, normalized spectra in normocapnia (black lines) and hypercapnia (red lines) are superimposed. A maximum of the T2 profiles appears around 20-30 Hz corresponding to the oscillation frequency observed on $S_{10}$ signal.

**Figure 4: Effects of hypercapnia on the CWT spectra of premetamorphic and postmetamorphic tadpoles**
Normalized CWT spectra for premetamorphic tadpoles (T1-T5) and postmetamorphic tadpoles (F1-F5) at pH=7.8 (black lines) and pH=7.4 (red lines). The values of the local maximal frequencies are indicated near the corresponding maxima of the profiles.

**Figure 5: Effects of hypercapnia on the amplitude profiles of the canonical buccal cycles**
Amplitude profiles of two canonical buccal cycles computed from the respective positive spike matrices on premetamorphic tadpoles (T1-T5) and postmetamorphic tadpoles (F1-F5) at pH=7.8 (black lines) and pH =7.4 (red lines). The profiles exhibit around 20-30 Hz oscillations as observed in the corresponding $S_{10}$ signals. The red horizontal segment indicates the interburst period lengthening induced by hypercapnia in both premetamorphic and postmetamorphic tadpoles.

**Figure 6: Contributions of the spike subpopulations on the buccal burst shape for T2**
The left column shows the spike maps at pH=7.8 (black dots) and pH=7.4 (red dots) with the corresponding mean number of spikes for each of the first four subpopulations (I to IV) into bins of 5 ms during a canonical cycle (black line for normocapnia and red line for hypercapnia).
The green lines delimit the canonical buccal burst (see Fig. 2). In the right column each amplitude profile is computed from its corresponding spike subpopulation, for pH=7.8 (black line) and pH=7.4 (red line).

**Figure 7: Contributions of the spike subpopulations on the buccal burst shape for F2**
Left and right columns exhibit respectively the number of spikes of each of the first four subpopulations (I to IV) and the corresponding amplitude profiles as in Fig. 6. Note that the number of buccal cycles is lower in hypercapnia than in normocapnia, for the same recording time, due to the marked decrease of the buccal frequency and the higher number of lung episodes which are eliminated from the analysis time.

**Figure 8: Overview of hypercapnia effect on the buccal and lung activities in neurograms of the postmetamorphic tadpoles (F1-F5).**
For each postmetamorphic tadpole, the nerve activity recorded during five minutes and filtered to get whole $S_{100}$ signals, i.e. with both buccal and lung activities, are represented according to a grey level scale. The time position of the lung bursts are indicated by vertical bars, black at pH=7.8 and red at pH=7.4. The grey levels indicate that the buccal activity is higher at pH=7.8 than at pH=7.4. The number of lung bursts is greater at pH=7.4 than at pH=7.8, but for F1.

**Figure 9: Example of lung bursts**
Left: typical lung bursts observed in F2 at pH=7.8 (black) and pH=7.4 (red).
In F2, all the bursts at pH=7.8 are preceded by a period of small amplitude activity, while such a pre-burst activity is absent at pH=7.4.
Right: mean amplitude of the lung burst and the pre-burst activity.
**Figure 10: Contributions of spike subpopulations II to V on the lung burst shape for F2**
Left and right columns exhibit respectively the number of spikes of each of the last four subpopulations (II to V) and the corresponding amplitude profiles as in Fig. 6 and 7. Note that the number of lung bursts is higher in hypercapnia than in normocapnia, for the same recording time.

FIGURE 1

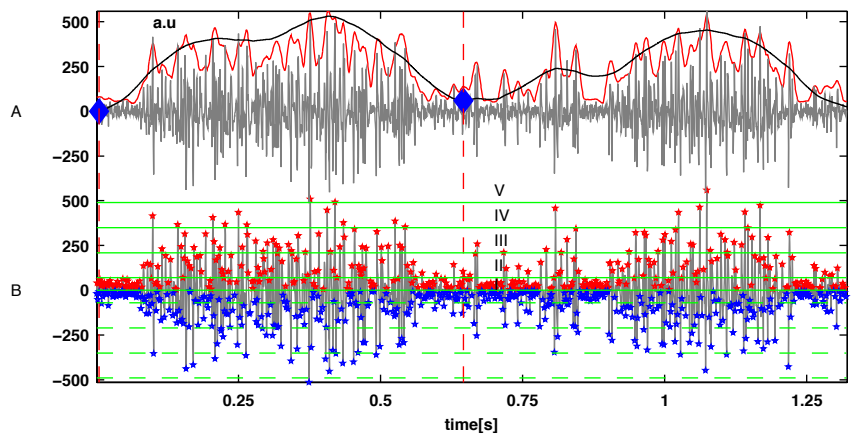

FIGURE 2

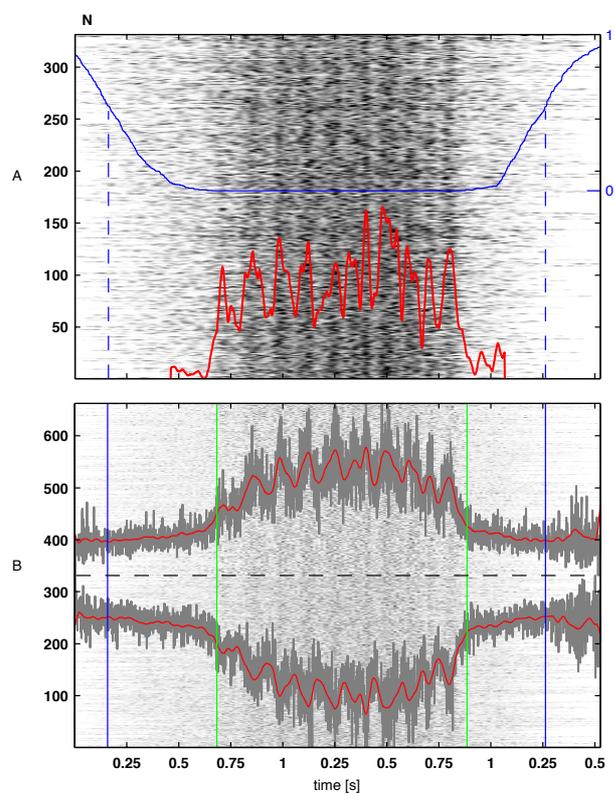

FIGURE 3

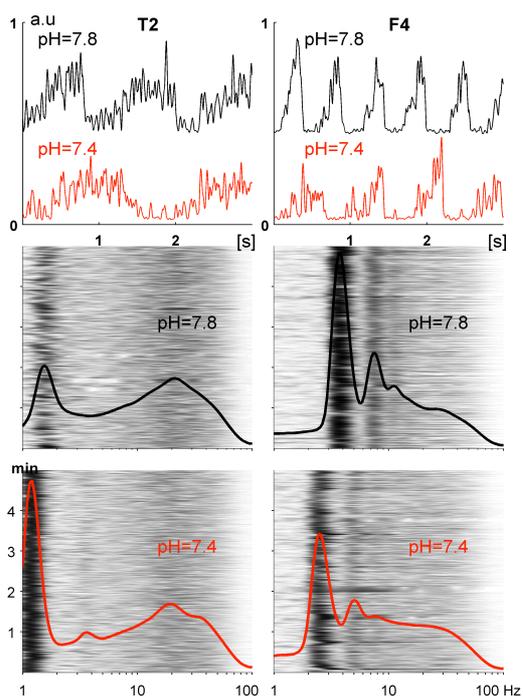

FIGURE 4

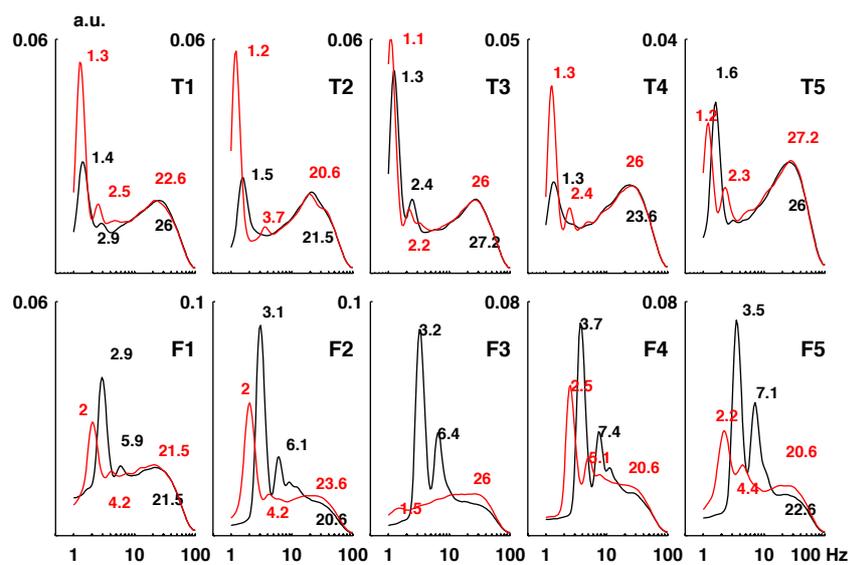

FIGURE 5

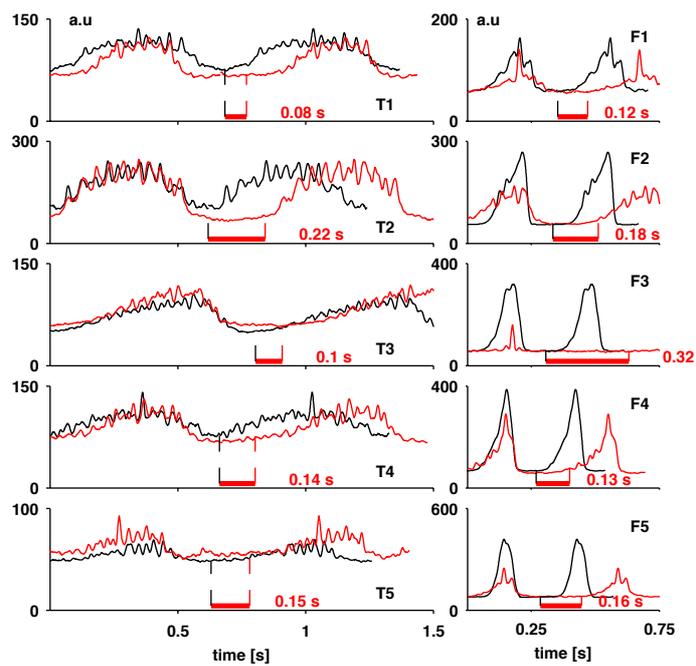

FIGURE 6

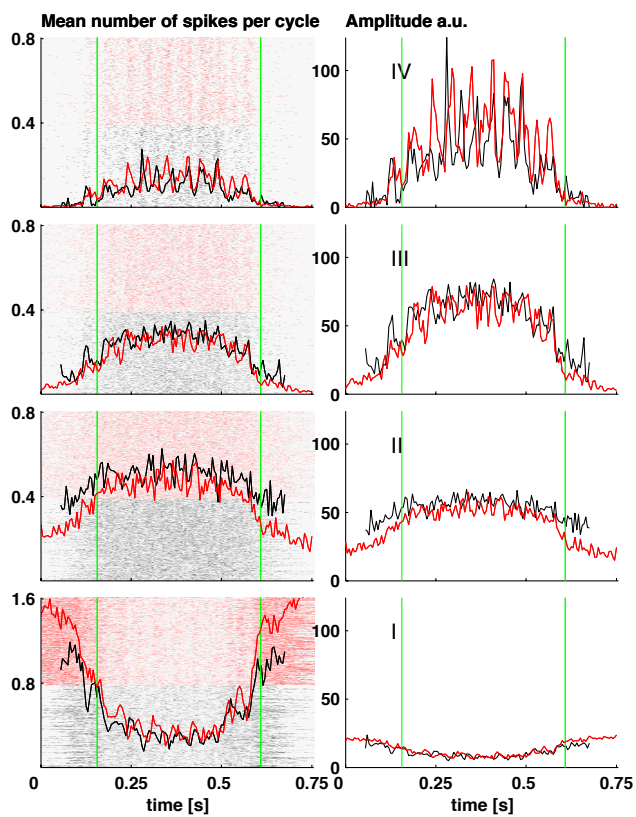

FIGURE 7

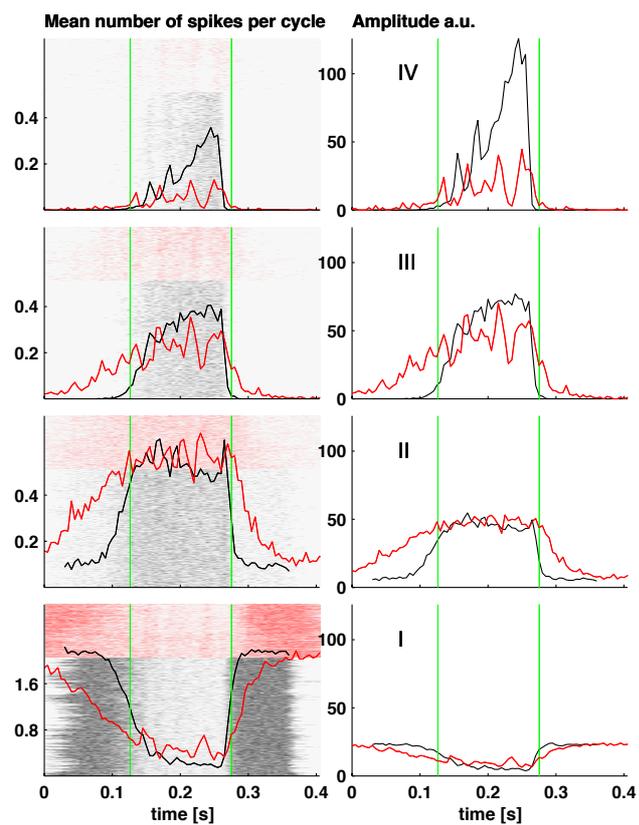

FIGURE 8

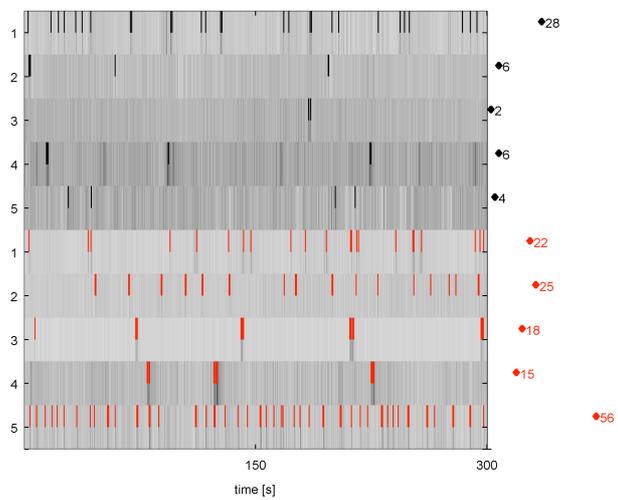

FIGURE 9

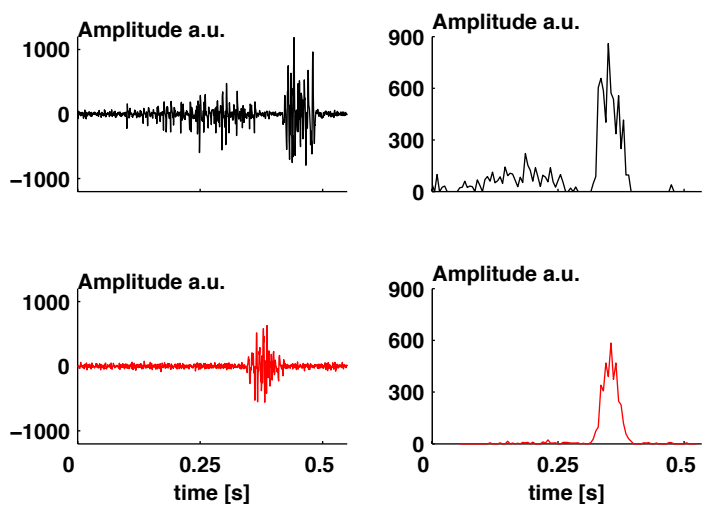

FIGURE 10

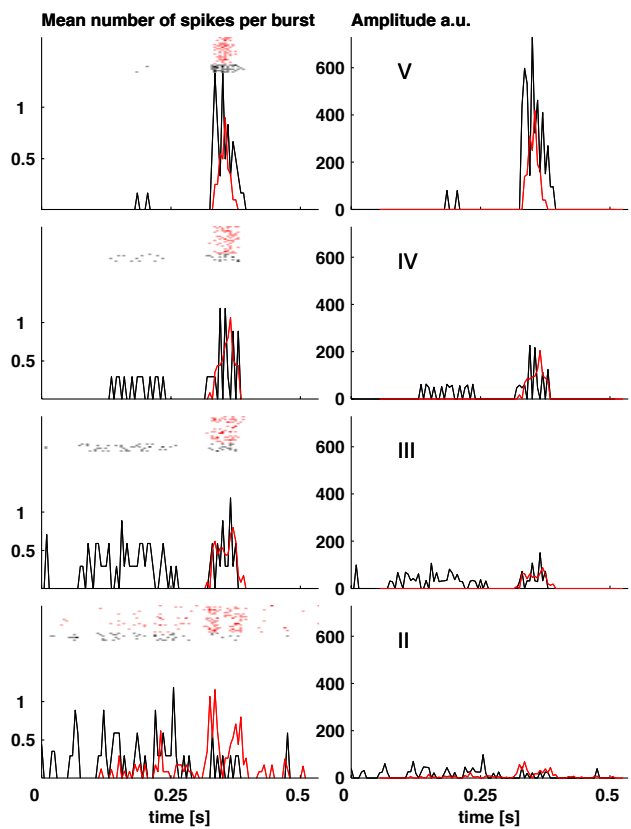